\documentstyle[12pt,epsf]{ioplppt}          % use this for preprint style

\begin{document}
%Title and Authors
\title{%
Anisotropic resistivity of 
the antiferromagnetic insulator Bi$_2$Sr$_2$ErCu$_2$O$_8$
}[Anisotropic resistivity of Bi$_2$Sr$_2$ErCu$_2$O$_8$]

\author{
T Kitajima, T Takayanagi, T Takemura and 
I Terasaki\ftnote{1}{To whom correspondence should be addressed.
E-mail address: terra@mn.waseda.ac.jp}
}

\address{
Department of Applied Physics, Waseda University,
Tokyo 169-8555, Japan
}

\begin{abstract}
The anisotropic resistivities of 
Bi$_2$Sr$_2$Ca$_{1-x}$Er$_x$Cu$_2$O$_8$ single crystals
were measured and analyzed from 4.2 to 500~K 
with special interest taken in the parent 
antiferromagnetic insulator with $x$=1.0.
Although the resistivity is semiconducting along both 
the in-plane and out-of-plane directions,
the temperature dependences are found to be significantly different.
As a result, the resistivity ratio for $x$=1.0 exhibits a broad maximum 
near room temperature.
The electric conduction in the parent antiferromagnetic insulators 
is different from that in other semiconductors,
and is as unconventional as that in high-temperature superconductors.
\end{abstract}

%\pacs{74.72.Hs, 74.25.Fy, 72.15.-v}

%---------------------------------------------------------------
%	Introduction
%---------------------------------------------------------------
\section{Introduction}
Anisotropic transport properties in the normal state
are one of the most striking features of
high-temperature superconductors (HTSC's) \cite{aniso}.
The metallic in-plane resistivity ($\rho_{ab}$)
accompanied by the non-metallic out-of-plane resistivity ($\rho_c$)
enhances $\rho_c/\rho_{ab}$ at low temperature ($T$) \cite{Y,La2},
whereas $\rho_c/\rho_{ab}$ is independent of $T$ for conventional metals.
The enhancement of $\rho_c/\rho_{ab}$
is often called `confinement' \cite{Anderson},
and can be a key to the elucidation of the mechanism of high-temperature 
superconductivity.
We have studied the anisotropic transport properties of 
slightly overdoped YBa$_2$Cu$_3$O$_y$ crystals \cite{terra1,terra2}.
Although their resistivities $\rho_c$ and $\rho_{ab}$ are both metallic,
the anisotropy is difficult to understand within the Fermi liquid theory.

The next question is that of whether
$\rho_c/\rho_{ab}$ is anomalous
for a parent antiferromagnetic (AF) insulator,
whose resistivities $\rho_c$ and $\rho_{ab}$ are semiconducting.
To our knowledge, very little investigation has been done on $\rho_c/\rho_{ab}$.
Thio {\it et al.} \cite{La1} have found 
that $\rho_c/\rho_{ab}$ for La$_2$CuO$_4$
{\it decreases} with decreasing $T$ below 200~K,
which is significantly incompatible with $\rho_c/\rho_{ab}$ for HTSC's.
Since it does not saturate near 200~K,
a higher-temperature measurement is needed.

For studying $\rho_c/\rho_{ab}$ over a wide temperature range,
Bi$_2$Sr$_2$Ca$_{1-x}R_x$Cu$_2$O$_8$ ($R$: rare-earth)
is most suitable for the following reasons:
(i) Oxygens for $x$=0 are chemically stable up to 600~K in air
where $\rho_{ab}$ remains ``$T$-linear'' \cite{T-linear}.
(ii) When $x$ reduces from 1 to 0,
the doping level varies from that of a parent AF insulator
to that of a (slightly) overdoped superconductor \cite{IMT}.
(iii) All of the Cu sites are equivalent,
and only the CuO$_2$ plane is responsible for the electric conduction.
Here we report on measurements and analyses of
Bi$_2$Sr$_2$Ca$_{1-x}$Er$_x$Cu$_2$O$_8$ single crystals
with $x$=0, 0.1, 0.5, and 1.0.
We have found that $\rho_c/\rho_{ab}$ for
a parent AF insulator ($x$=1.0) is quite unique;
it increases with $T$ below 100 K, 
takes a broad maximum near 300 K,
and decreases above room temperature.
This obviously indicates that a parent AF insulator exhibits
a quite different conduction mechanism from conventional semiconductors.

\begin{table}[t]
\caption{%
Characterization of Bi$_2$Sr$_2$Ca$_{1-x}$Er$_x$Cu$_2$O$_8$
single crystals.
Note that the actual composition ratio is represented
by setting Cu = 2.
}
\begin{indented}
\item[]\begin{tabular}{llcc}
Nominal   & Actual Composition         & Size                     & $c$-axis  \\
$x$       & Bi : Sr : Ca : Er : Cu   & (mm$^3$)                 & (\AA)   \\
\hline
0         &1.9 : 1.9 : 1.2 : 0   : 2 &0.6$\times$1$\times$0.02  &30.85    \\
0.1       &1.6 : 1.8 : 1.2 : 0.1 : 2 &1$\times$1.2$\times$0.02  &30.90    \\
0.5       &1.6 : 1.9 : 1.0 : 0.5 : 2 &1$\times$1$\times$0.004   &30.32    \\
1.0       &2.0 : 2.1 : 0   : 0.6 : 2 &1$\times$1.2$\times$0.004 &30.33    \\
\end{tabular}
\end{indented}
\end{table}

%---------------------------------------------------------------
%	Experimental
%---------------------------------------------------------------
\section{Experimental}
Single crystals of Bi$_2$Sr$_2$Ca$_{1-x}$Er$_x$Cu$_2$O$_8$
were grown by a self-flux method \cite{preparation}.
Powders of Bi$_2$O$_3$, SrCO$_3$, CuO, CaCO$_3$, and Er$_2$O$_3$
of 99.9~\% purity were mixed,
well ground in an Al$_2$O$_3$ crucible,
heated at 900$^{\circ}$C [1020$^{\circ}$C] for 10~h,
and slowly cooled down to 760$^{\circ}$C [820$^{\circ}$C]
by 2$^{\circ}$C/h for $x$=0 [$x\neq$0].
Since the single crystals were very thin along the $c$ axis,
the thickness was measured with a scanning electron microscope (SEM).
The actual compositions were measured through 
inductively coupled plasma emission spectroscopy.
The x-ray diffraction pattern showed no trace of impurities,
and the $c$-axis lattice parameter for $x$=0 was evaluated to be 30.85~\AA,
which agrees with the value in the literature \cite{preparation,x-ray}.
The measured compositions, sizes, and $c$-axis lattice parameters
are listed in table 1.
We should note that crystals grown by a flux method are produced 
with little stress, owing to the slow cooling rate near thermal equilibrium.
In fact, we did not observe any serious cracks 
in the SEM images of our samples.
In order to examine the influence of inhomogeneity and disorder 
on the resistivity, 
we made measurements for more than 30 samples 
including ones grown from different batches.
The measured resistivities were reproducible enough to warrant
the discussion in this paper.

The resistivity was measured with a $dc$ current $I$
in a four-probe configuration
along the in-plane direction ($I\perp c$), 
and in a ring configuration along the out-of-plane
direction ($I\parallel c$).
We used two measurement systems below and above room temperature.
From 4.2 to 300 K, the samples were slowly (100 K/h) cooled 
in a liquid-He cryostat,
where $T$ was monitored through a cernox resistance thermometer.
Above 300 K, the samples are slowly (50-100 K/h)
heated in air in a cylinder furnace
with a Pt resistance thermometer.

\begin{figure}[t]
\centerline{\epsfxsize=7cm 
\epsfbox{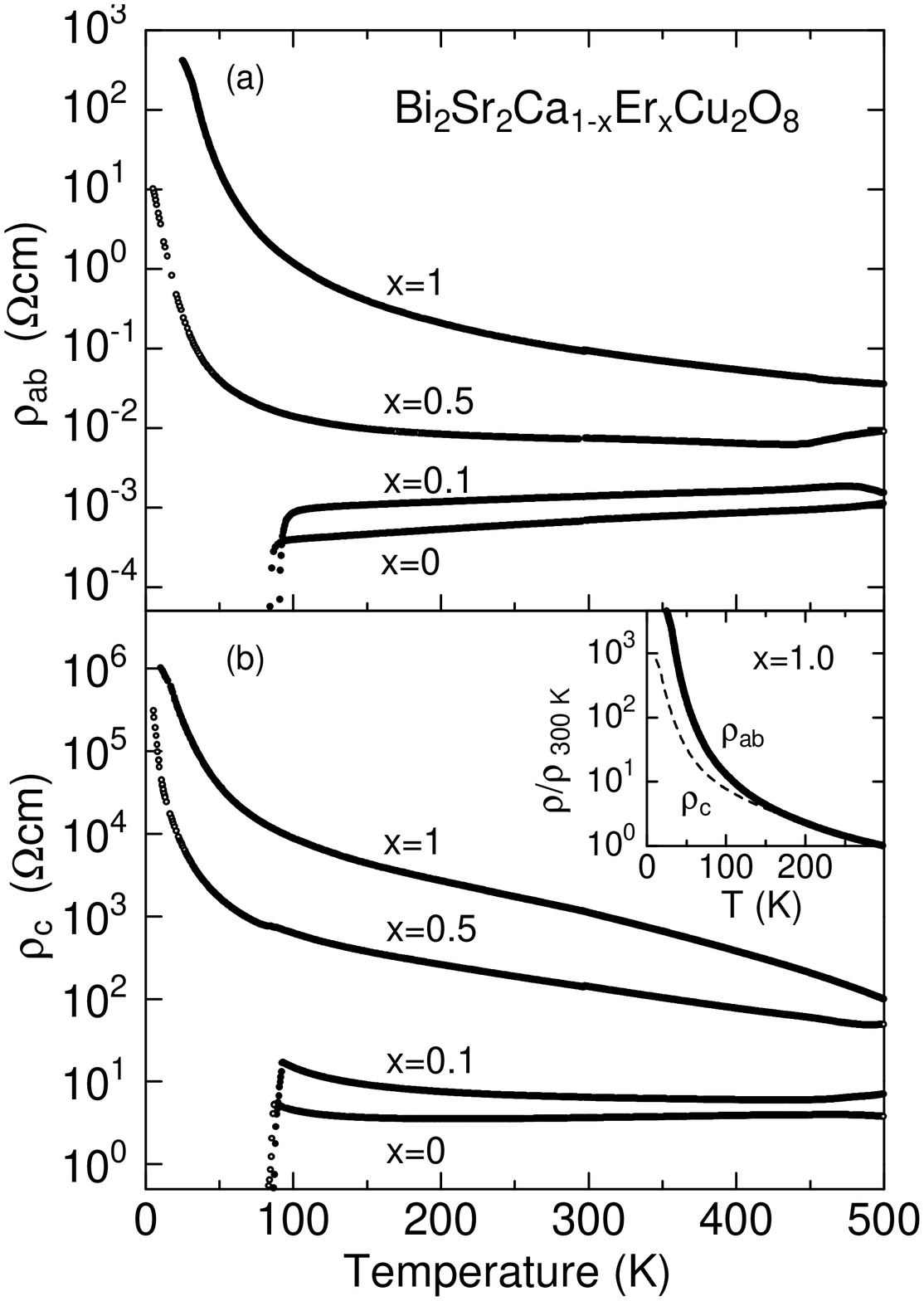}
}
\caption{%
(a) The in-plane resistivity $\rho_{ab}$ and
(b) out-of-plane resistivity $\rho_c$
of single crystals of Bi$_2$Sr$_2$Ca$_{1-x}$Er$_x$Cu$_2$O$_8$.
$\rho_{ab}$ and $\rho_c$ for $x$=1.0 
normalized at 300 K are plotted as functions of temperature 
in the inset.}
\end{figure}

%---------------------------------------------------------------
%	Results and Discussion
%---------------------------------------------------------------
\section{Results and discussion}
Figures 1(a) and 1(b) show $\rho_{ab}$ and $\rho_c$ of
Bi$_2$Sr$_2$Ca$_{1-x}$Er$_x$Cu$_2$O$_8$ single crystals, respectively.
The magnitudes of $\rho_{ab}$ and $\rho_c$ increase with $x$,
showing that the hole concentration is reduced by the Er substitution.
As is seen in the literature,
$\rho_c$ is four or five orders of magnitude
lager than $\rho_{ab}$ for all the samples.
For superconducting samples ($x$=0 and 0.1),
metallic $\rho_{ab}$ and semiconducting $\rho_c$ are observed above $T_c$.
Reflecting the slightly overdoped nature of $x$=0,
$T_c$ ($\sim$84~K) for the $x$=0 sample 
is lower than $T_c$ ($\sim$87~K) for $x$=0.1.
These results attest to the quality of our crystals.

Both $\rho_{ab}$ and $\rho_c$ for $x$=1.0 are semiconducting,
but they exhibit different $T$ dependences.
Above room temperature,
where $\rho_{ab}$ decreases gradually in comparison with $\rho_c$,
$\rho_c/\rho_{ab}$ decreases with increasing $T$.
On the other hand, $\rho_{ab}$ becomes insulating more rapidly 
than $\rho_c$, as shown in the inset of figure 1
where the resistivities are normalized at 300 K.
Thus $\rho_c/\rho_{ab}$ decreases with decreasing $T$ below 300 K.
These results are not understandable on the basis of conventional theories.
In the framework of a band picture, anisotropy is 
mainly determined by effective masses,
implying that the $T$ dependence of $\rho$ is independent of 
the direction.
In the case of a hole doped in the AF background, 
the situations are nearly the same.
In fact, a nearly $T$-independent $\rho_c/\rho_{ab}$ has been  observed
for La$_2$NiO$_4$ \cite{LaNi}
and Bi$_2M_3$Co$_2$O$_9$ ($M$=Ca, Sr and Ba) \cite{BiCo}.

\begin{figure}[t]
\centerline{\epsfxsize=7cm 
\epsfbox{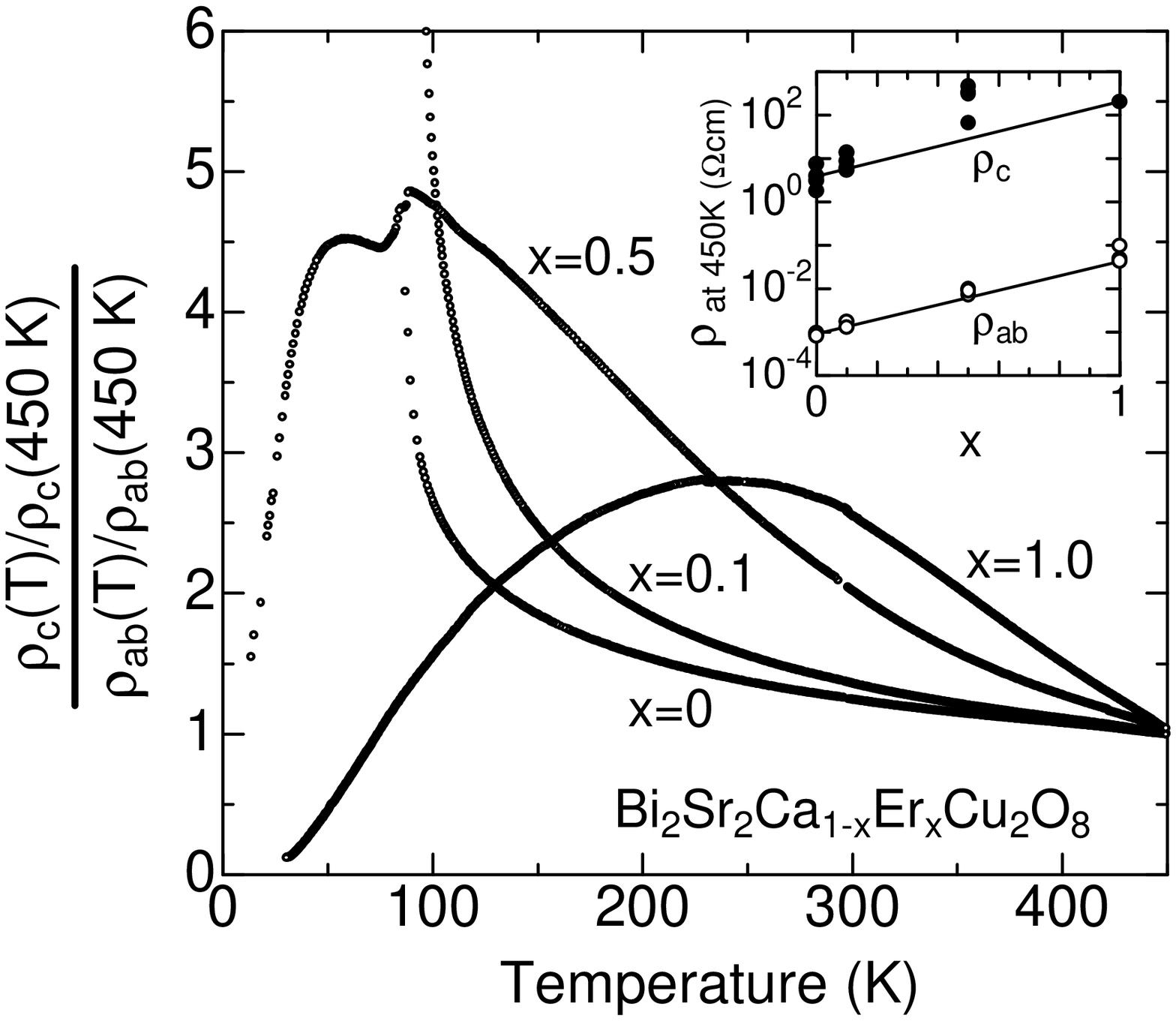}
}
\caption{%
The anisotropic resistivity ratios $\rho_c/\rho_{ab}$
of Bi$_2$Sr$_2$Ca$_{1-x}$Er$_x$Cu$_2$O$_8$ single crystals
normalized at 450~K.
Inset: The magnitudes of $\rho_{ab}$ and $\rho_c$ at 450~K are plotted
as functions of $x$.
Note that each symbol represents the resistivity of a different sample
measured to check the reproducibility (see text).
}
\end{figure}

The magnitude of $\rho_c/\rho_{ab}$ for a parent AF insulator
is much more difficult to evaluate than that for a superconductor,
in that it is an exponentially varying quantity divided by another 
exponentially varying quantity.
Since we are interested in the $T$ dependence of $\rho_c/\rho_{ab}$,
we normalize $\rho_c/\rho_{ab}$ at 450~K in figure 2.
As for the magnitude, we show $\rho_{ab}$ and $\rho_c$
in the inset of figure 2,
in which each symbol corresponds to a different sample.
From the inset one can see that the magnitude of $\rho_c/\rho_{ab}$ at 450 K 
is nearly independent of $x$.
Accordingly the normalization at 450 K
will not adversely affect the discussion below.

We would like to point out three features in figure 2.
First, $\rho_c/\rho_{ab}$ changes smoothly with $x$ 
above room temperature;
It increases with decreasing $T$, and the $T$ dependence is
steeper for larger $x$ (smaller hole concentration).
If one looked at $\rho_c/\rho_{ab}$ only above room temperature,
one could not distinguish a parent AF insulator from a superconductor. 
Thus we may say that the holes are
confined in a parent AF insulator as well as in HTSC.
In this context  the former is as unconventional as the latter.
Secondly, $\rho_c/\rho_{ab}$ for $x$=1.0 and 0.5 
decreases with decreasing $T$ below 100 K, 
which is consistent with $\rho_c/\rho_{ab}$ of La$_2$CuO$_4$ \cite{La1}.
The decrease of $\rho_c/\rho_{ab}$ as $T\to$0 could be understood
from the three-dimensional (3D) nature of the localization \cite{Localize}.
Thirdly, $\rho_c/\rho_{ab}$ for $x$=1.0 and 0.5
takes a maximum at a certain temperature $T_{\rm max}$,
which is very close to the N\'eel temperature $T_N$.
(For $x$=0.5, a tiny fraction of a superconducting phase causes a small 
drop of resistivity near 90 K, which obscures the position of $T_{\rm max}$.)

\begin{figure}[t]
\centerline{\epsfxsize=7cm 
\epsfbox{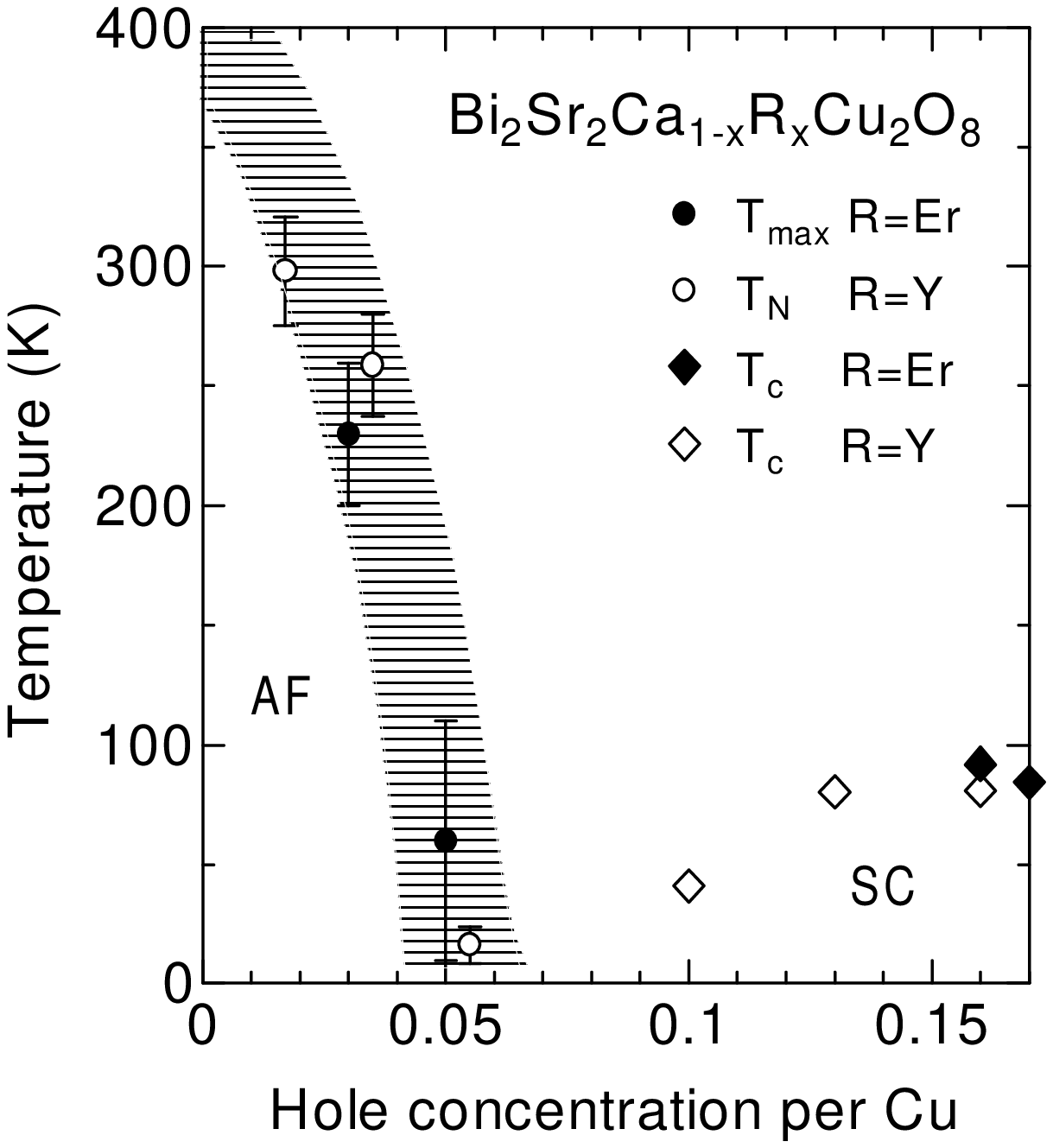}
}
\caption{%
The phase diagram of Bi$_2$Sr$_2$Ca$_{1-x}R_x$Cu$_2$O$_8$ ($R$=Er and Y).
AF and SC represent the antiferromagnetic order and the superconducting phase,
respectively.
$T_{\rm max}$ (the temperature at which $\rho_c/\rho_{ab}$ takes a maximum),
$T_N$ (the N\'eel temperature), and $T_c$ (the superconducting transition
 temperature) are plotted as a function of hole concentration.
$T_N$ and $T_c$ for $R$=Y are taken 
from Refs. \protect{\cite{T_N}} and \protect{\cite{Y-S}}.
The hole concentration is estimated from the room-temperature thermopower,
as is proposed in Ref. \protect{\cite{S-p}}.
The error bars in $T_{\rm max}$ represent the variation from sample to sample.
}
\end{figure}

The localized $f$ spins of Er$^{3+}$
make it difficult to measure $T_N$ of Cu$^{2+}$ for $x$=1.0 and 0.5.
Instead, we will compare
$T_N$ of Bi$_2$Sr$_2$Ca$_{1-x}$Y$_x$Cu$_2$O$_8$ \cite{T_N},
and we plot $T_c$, $T_N$ and $T_{\rm max}$ 
in the electronic phase diagram of
Bi$_2$Sr$_2$Ca$_{1-x}R_x$Cu$_2$O$_8$ in figure 3.
We estimated the hole concentration
using an empirical relation to the thermopower \cite{S-p},
which we measured with the same samples for $R$=Er (not shown here), 
and used Ref. \cite{Y-S} for $R$=Y.
$T_{\rm max}$  is found to lie around the AF boundary.
Since no structural transitions
and no phase separations are reported for
Bi$_2$Sr$_2$Ca$_{1-x}R_x$Cu$_2$O$_8$ \cite{IMT},
it would be natural to relate 
$T_{\rm max}$ to the AF transition.

The confinement behavior above $T_N$ 
favors some theories based on the
two-dimensional (2D) spin fluctuation,
which exists in parent AF insulators above $T_N$ \cite{Shirane}
and in HTSC's above $T_c$ as well \cite{Yamada}.
We therefore propose 
that holes are confined in a CuO$_2$ plane
in the presence of the 2D spin fluctuation,
{\it which occurs irrespective of doping levels}.
As the 3D AF order grows below $T_N$, 
the confinement becomes less effective.
A recent numerical analysis of the bilayer $t-J$ model 
has also led to the assertion that $\rho_c$ is substantially modified
in the presence of the 2D spin fluctuation \cite{Eder}.
We further note that a similar case is seen for a layered ferromagnet
La$_{2-x}$Sr$_{1+x}$Mn$_2$O$_7$ \cite{kimura}.
For 100 K $<T<$ 250 K, this compound is in a 2D ferromagnetic state, 
and exhibits a non-metallic $\rho_c$ together with a metallic $\rho_{ab}$.
Once the 3D ferromagnetic order appears below 100 K,
$\rho_c$ becomes metallic to behave in a 3D-like manner.
We believe that the out-of-plane conduction in parent AF insulators 
includes essentially the same physics as for La$_{2-x}$Sr$_{1+x}$Mn$_2$O$_7$;
the only difference is as regards whether the material
is an antiferromagnetic insulator or a ferromagnetic metal.

%---------------------------------------------------------------
%	Summary
%---------------------------------------------------------------
\section{Summary}
We prepared Bi$_2$Sr$_2$Ca$_{1-x}$Er$_x$Cu$_2$O$_8$,
single crystals for $x$=0, 0.1, 0.5 and 1.0
and measured the in-plane and out-of-plane resistivities 
($\rho_{ab}$ and $\rho_c$) from 4.2 to 500~K.
The present study has revealed that $\rho_c/\rho_{ab}$
for a parent antiferromagnetic insulator 
($x=1.0$) strongly depends on temperature,
and that the enhancement of $\rho_c/\rho_{ab}$ 
with decreasing $T$ is observed above room temperature.  
In this sense, parent antiferromagnetic insulators
are as unconventional as high-temperature superconductors.
Their ratios $\rho_c/\rho_{ab}$ take maxima at a certain temperature
near the N\'eel temperature,
and we propose that the confinement in the CuO$_2$ plane is operative
in the two-dimensional spin-fluctuation regime
regardless of the doping level.

%---------------------------------------------------------------
%	Acknowledgments
%---------------------------------------------------------------
\section*{Acknowledgments}
The authors would like to thank T Itoh, T Kawata, K Takahata and Y Iguchi
for collaboration.
They also wish to express their appreciation to S  Kurihara and S  Saito 
for fruitful discussions and valuable comments.
One of the authors (I T) is indebted to
S Tajima for the collaboration at a preliminary stage of this work.
This work was partially supported by Waseda University Grant
for Special Research Projects (97A-565, 98A-618).

%---------------------------------------------------------------
%	References
%---------------------------------------------------------------

\end{document}